\documentstyle{mn2e}
\input psfig

\def\solmas{{M$_\odot$}}
\def\simless{\mathbin{\lower 3pt\hbox
   {$\rlap{\raise 5pt\hbox{$\char'074$}}\mathchar"7218$}}}   
\def\simgreat{\mathbin{\lower 3pt\hbox
   {$\rlap{\raise 5pt\hbox{$\char'076$}}\mathchar"7218$}}}   
\def\etal{{\rm et al.}}
\def\mj {M_{\rm Jeans}}

\def\solmas{{M$_\odot$}}
\def\solm{{M_\odot}}

\def\au{AU}

\def\tff {t_{\rm ff}}

\def\apj{{ApJ}}

\def\mnras{{MNRAS}}

\def\be{\begin{equation}}
\def\ee{\end{equation}}

  \makeatletter
  \ifx\CUP@mtlplain@loaded\undefined  
  \makeatother  
    
  \else
    
  \fi

\loadboldmathitalic 
\loadboldgreek      
  
\makeatletter
\ifx\CUP@mtlplain@loaded\undefined  
  \newfont\bit{cmbxti10 at 9pt}
\else
  \newfont\bit{mtbxti10 at 9pt}
\fi
\makeatother

\def\LaTeX{L\kern-.36em\raise.3ex\hbox{a}\kern-.15em
    T\kern-.1667em\lower.7ex\hbox{E}\kern-.125emX}

\newcommand{\gsim}{\mathrel{\hbox{\rlap{\lower.55ex \hbox {$\sim$}}
                   \kern-.3em \raise.4ex \hbox{$>$}}}}
\newcommand{\lsim}{\mathrel{\hbox{\rlap{\lower.55ex \hbox {$\sim$}}
                   \kern-.3em \raise.4ex \hbox{$<$}}}}

\title[The knee of the IMF] {The Jeans mass and the origin of the knee in the IMF}
\author[I. A. Bonnell \etal]
  {I. A. Bonnell$^1$\thanks{E-mail: iab1@st-and.ac.uk},  C. J. Clarke$^2$ \& M. R. Bate$^3$ \\
$^1$ School of Physics and
  Astronomy, University of St Andrews, North Haugh, St Andrews, Fife,
  KY16 9SS. \\
$^2$ Institute of Astronomy, Madingley Road, Cambridge, CB3 0HA \\ 
$^3$  School of Physics, University of Exeter, Stocker Road, Exeter, EX4 4QL \\}

\date{\today}

\begin{document}

\maketitle

\begin{abstract}

We use numerical simulations of the fragmentation of a 1000 \solmas\ molecular cloud and the formation 
of a stellar cluster to study how the initial conditions for star formation affect the resulting initial mass 
function (IMF). In particular, we are interested in the relation between the
thermal Jeans mass in a cloud and the {\sl knee} of the initial mass function, i.e.  the mass separating
 the region with a flat IMF slope from that typified by a steeper, Salpeter-like, slope.
In three isothermal simulations  with $\mj=1 \solm$, $\mj =2\solm$ and $\mj=5\solm$,  the number of stars
formed, at comparable dynamical times,  scales roughly
with the number of initial Jeans masses in the cloud. The mean stellar
mass also increases (though less than linearly) with the
initial Jeans mass in the cloud. It is found that the IMF in each
case displays a prominent knee, located roughly at the mass scale of
the initial Jeans mass. Thus clouds with higher initial Jeans masses
produce  IMFs which are shallow to higher masses. This implies that a universal IMF requires
a physical mechanism that sets the Jeans mass to be near 1 \solmas. Simulations including a barotropic
equation of state as suggested by Larson, with cooling at low densities followed by gentle heating at higher
densities, are able to produce realistic IMFs with the knee located at $\approx 1 \solm$, 
even with  an initial $\mj=5\solm$. 
We therefore suggest that the observed universality of the IMF in the local
Universe does not require any fine tuning of the initial conditions
in star forming clouds but is instead imprinted by details 
of the cooling physics of the collapsing
gas.

\end{abstract}


\section{Introduction}

One of the goals of studies of star formation is to understand the origin of the initial
mass function (IMF). Since Salpeter's initial study (Salpeter 1955), we have known that
a decreasing number of stars form with increasing mass. The form of this mass function,
\be
dN \propto m^{-\gamma} dm
\ee
with $\gamma=2.35$, 
has proven to be remarkably robust for intermediate and high-mass stars (Kroupa 2002).
More recently,
we have seen that the distribution of lower-mass stars is less steep (Miller \& Scalo 1979; Scalo 1986; Kroupa \etal\ 1993), $\gamma=1.5$, 
eventually becoming nearly flat for brown dwarfs (Allen \etal\ 2005; Kroupa 2002).
This {\sl knee}  in the slope of the mass distribution suggests that there exists a change in the physical
process which determines the  stellar masses (Elmegreen 2004) for low and high-masses. The  location of this
knee at $\approx 1 \solm$ is found to be fairly consistent across the Galactic stellar populations (Kroupa 2002; Munch \etal 2002; Chabrier 2003)
and thus provides a useful reference point for models of star formation and the initial mass function.

There have been many theories developed to explain the origin of the initial mass function. These
theories are based upon  physical processes such as fragmentation (Zinnecker 1984; Larson 1985;
Klessen, Burkert \& Bate 1998; Klessen \& Burkert 2001; Klessen 2001), turbulence (Padoan \& Nordlund 2002), coagulation (Silk \& Takahashi 1979; Murray \& Lin 1996), accretion  (Zinnecker 1982; Bonnell \etal\ 1997; Klessen \& Burkert 2000; Bonnell \etal\ 2001b; Basu \& Jones 2004; Bate \& Bonnell 2005),
stellar mergers (Bonnell, Bate \& Zinnecker 1998; Bonnell \& Bate 2002) and feedback (Silk 1995),
or a combination  of multiple processes (Adams \& Fatuzzo 1996). In addition to being able to reproduce the slope
of the IMF, any theory needs to be able to explain the characteristic stellar mass. This characteristic mass occurs due to the break in the IMF and it is therefore crucial to understand why there is a {\sl knee}
in the IMF. It is also crucial that these theories establish secondary indicators, such as the
initial mass segregation in stellar clusters (Bonnell \& Davies  1998; Bonnell \etal~2001a) in order to assess
their relevance.

Numerical simulations of the collapse and fragmentation of turbulent molecular clouds 
are now capable of following  the formation of sufficient numbers of stars to resolve the IMF
from the low-mass brown dwarfs (Bate, Bonnell \& Bromm 2003) to high-mass O stars (Bonnell, Bate \& Vine 2003).  These simulations produce nearly flat IMFs ($\gamma \approx 1$) for low-mass
objects (Bate \etal\ 2003; Bate \& Bonnell 2005), rolling over to $\gamma=1.5$ near $m\simless 1 \solm$ and then to $\gamma\approx 2.5$
for higher-mass stars (Bonnell \etal\ 2001b; Bonnell \etal\ 2003). 

In order for fragmentation to occur, significant structure is needed in the molecular clouds.
Turbulence can generate significant structure due to shocks driven into the gas (e.g. Larson 1981; Padoan \etal 2001). A detailed analysis of the pre-star formation evolution
shows that the turbulent-driven structure is generally unbound and needs to grow through coagulation
and accretion onto the clumps before gravitational collapse occurs (Clark \& Bonnell 2005). Significantly, the mass at which
the clumps become gravitationally bound is that of the thermal Jeans mass in the unperturbed cloud.
The Jeans mass is given by
\be
\mj = \left(\frac{5 R_g T}{2 G \mu}\right)^{3/2} \left(\frac{4}{3} \pi
\rho \right)^{-1/2},  
\ee where $\rho$ is the gas density, $T$ the temperature, $R_g$ is the gas constant, $G$ is the gravitational constant and $\mu$ is 
the mean molecular weight of the gas.

Larson (1985, 2005) has suggested that a change in the gas cooling
with densities may provide an important determinant of the Jeans mass at the point of fragmentation. This change occurs as the cooling is dominated at lower densities by atomic and molecular line emission whereas at higher densities the gas is coupled to the dust and it is
dust cooling that dominates. The resulting equation of state is suggested to involve cooling at lower densities followed by a gentle heating at higher densities. Jappsen \etal (2005) have used
such an equation of state in numerical simulations and shown that it can set the mass scale
for star formation. 

It is therefore likely that the thermal Jeans mass plays an important role in determining the characteristic
stellar mass (Larson 2005).  Fragmentation, and ejections, most probably play a key role in determining the low-mass end of the IMF
(Larson 1985; Bate \etal 2003; Bate \& Bonnell 2005) whereas competitive accretion or other coagulation process dominates the formation of intermediate
and massive stars (Larson 1982;  Zinnecker 1982;  Bonnell \etal\ 2001a; Bonnell \& Bate 2002; Bonnell, Vine \& Bate 2004; Bonnell \& Bate 2005). In this paper we use numerical simulations to investigate the direct relation
between the thermal Jeans mass and the {\sl knee} of the IMF.

\begin{figure*}
\centerline{\psfig{figure=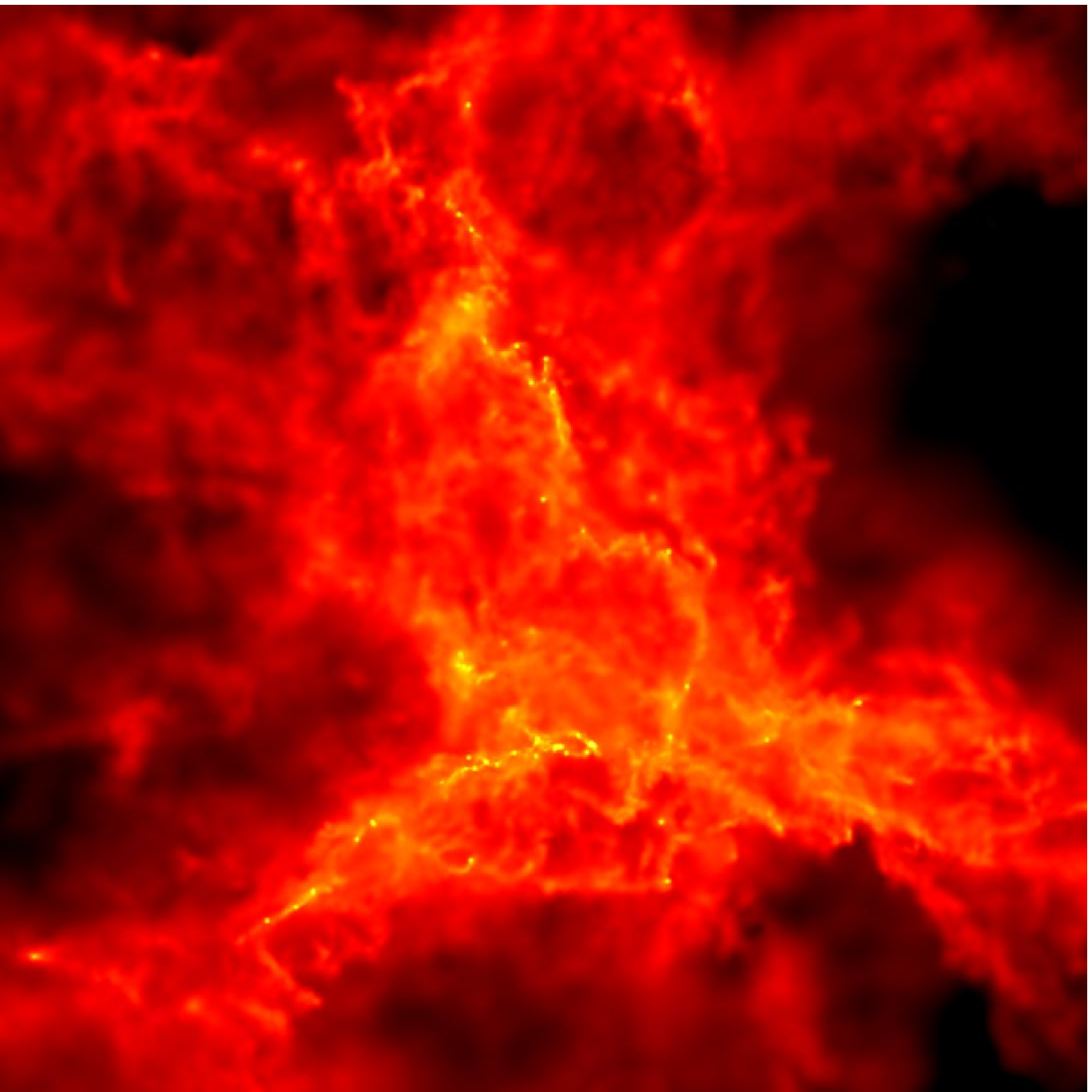,width=5.5truecm,height=5.5truecm}\psfig{figure=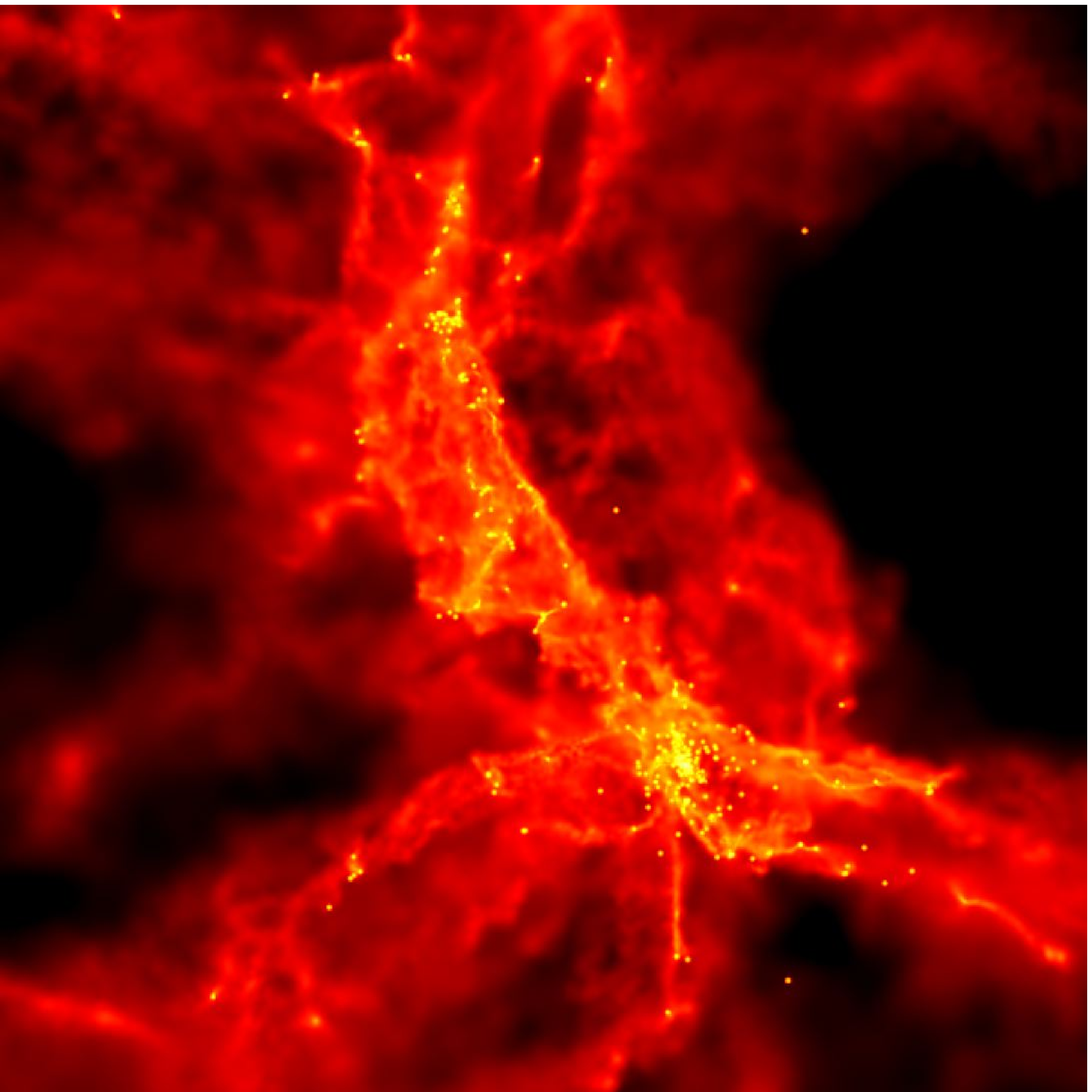,width=5.5truecm,height=5.5truecm}\psfig{figure=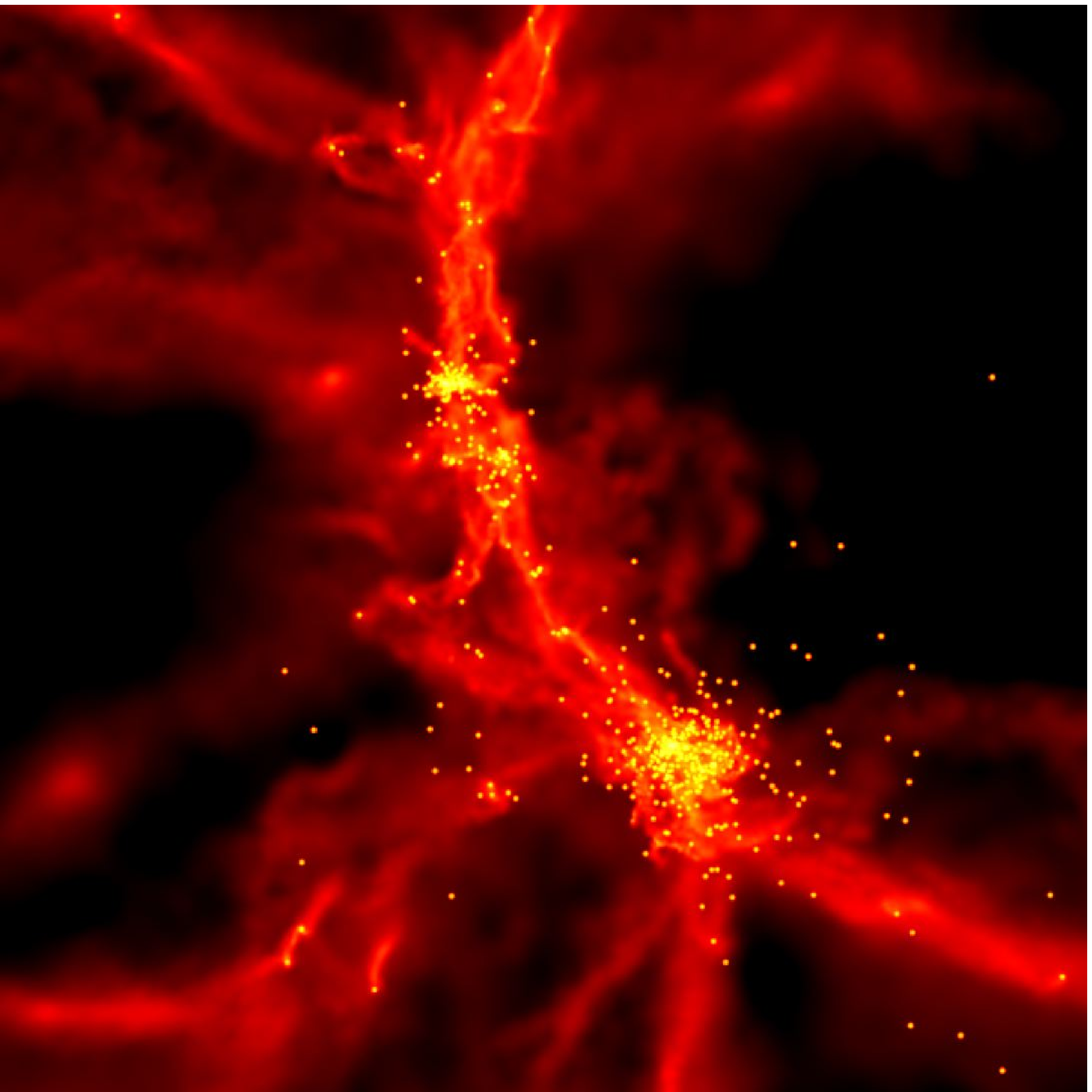,width=5.5truecm,height=5.5truecm}}
\caption{\label{cluform}  The evolution of the cluster forming  in simulation A with $\mj=1 \solm$ is shown at $t=0.91 \tff$, $t=1.34 \tff$, and $t=1.72 \tff$. The column density is plotted from a minimum value of $0.0075$ to a maximum value of 150 g cm$^{-2}$.}
\end{figure*}

\section{Calculations}

\begin{table} \caption{\label{simulations} Status of simulations at $t=1.5 \tff$. The radii of the clouds is in pc and the mases are in \solmas.}
\begin{tabular}{c|c|c|c|c|c|c|c}
Sim & $M_{\rm J}$ & EoS & Radius & N$_{\rm stars}$ & M$_{\rm tot}$   &M$_{\rm med}$ & M$_{\rm av}$\\
\hline
\hline
A & 1 & Iso & 0.5  &  399  & 330 &  $ 0.35$ & $ 0.82$  \\
B & 2 & Iso & 0.8  &  231  & 316  &  $ 0.58$  & $ 1.37$  \\
C & 5 & Iso & 1.5  &  91   & 195  &  $ 0.63$  & $ 2.15$  \\
D & 5 & Lar & 0.9 & 334 &  333  & $0.47$   & $1.00$ \\
E & 5 & Iso & 0.9 & 119 &  253  & $0.60$   & $2.13$ \\
\hline
\end{tabular} \end{table}

We use Smoothed Particle Hydrodynamics (SPH)
(Monaghan~1992, Benz \etal~1990) simulations to follow the  fragmentation
of 1000 \solmas\ molecular clouds and the formation of sufficient numbers of stars to
resolve the form of the IMF.   The goal is to investigate the dependency of the knee of the IMF 
on the Jeans mass in the cloud. We perform three isothermal simulations varying the
initial size of the cloud while maintaining the temperature at 10 K. The gas is assumed
to remain isothermal in these simulations. The cloud sizes
are set (see Table~1) such that the initial Jeans mass of the cloud is either $1\solm$,  $2\solm$, 
or $5\solm$.   We also report on one non-isothermal simulation where we use a barotropic
equation of state to mimic the transition from line to dust cooling invoked by Larson (2005).
This equation of state has an initial cooling of the form
\be
T = T_c  \left({\frac{\rho}{\rho_c}}\right)^{-0.25},      \rho < \rho_c
\ee
followed by a gentle heating
\be
T = T_c  \left({\frac{\rho}{\rho_c}}\right)^{0.1},       \rho > \rho_c
\ee
once the gas is well coupled to the dust. The critical temperature and density
was chosen to be $T_c= 4.9 K$ at $\rho_c = 3 \times 10^{-18}$ g cm$^{-3}$. This simulation started
from a density of $\rho=2.2 \times10^{-20}$ g cm$^{-3}$
and a temperature of 16. K, and thus a Jeans mass of $5.\solm$. In order to facilitate
comparison, we have also performed an isothermal simulation from the same initial conditions.

Star formation is modeled by the inclusion of sink-particles (Bate, Bonnell \& Price 1995)
that interact only through self-gravity and through gas accretion. Sink-particle creation occurs when 
dense clumps   of gas have  $\rho \simgreat 1.5 \times 10^{-15}$ g cm$^{-3}$, are
self-gravitating, and are contained   in a region such that the SPH smoothing lengths are smaller
than the  'sink radius' of  200 \au. Gas particles are accreted if they fall within a sink-radius (200 \au) of a sink-particle and are bound to it.
In the case of overlapping sink-radii, the gas particle is accreted by the sink-particle to which
it is most bound. The gravitational forces between sink-particles are smoothed at $160$ \au\ 
using the SPH kernel. 
 The simulations contain $5 \times 10^5$ particles such that the minimum resolvable mass
 is $0.15 \solm$ (Bate \& Burkert 1997; Bate \etal 2003). We therefore cannot resolve any star formation below this  value although 
 sink-particles commonly form with initial masses typically half this value,
 corresponding to the number of neighbours in one SPH smoothing length. These 
 'stars' quickly accrete up to the minimum resolvable mass.
The code has variable smoothing
lengths in time and in space and solves for the self-gravity of the
gas and stars using a tree-code (Benz \etal~1990). 
 The simulations were
carried out on the United Kingdom's Astrophysical Fluids Facility
(UKAFF), a 128 CPU SGI Origin 3000 supercomputer.


\begin{figure}
\centerline{\psfig{figure=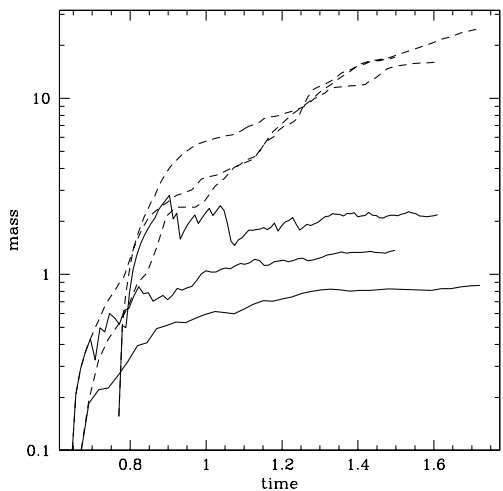,width=8.truecm,height=8.truecm}}
\caption{\label{massevol} The mean (solid lines) and maximum (dashed lines) stellar masses for the three isothermal simulations are plotted
as a function of the free-fall time.  The mean stellar masses are, from top to bottom, for initial
Jeans masses of 5, 2 and 1 \solmas.  The mean stellar masses increase with the
Jeans mass. The maximum stellar masses are basically indistinguishable
and reflect solely the total mass being accreted into the forming stellar clusters. }
\end{figure}

\begin{figure*}
\centerline{\psfig{figure=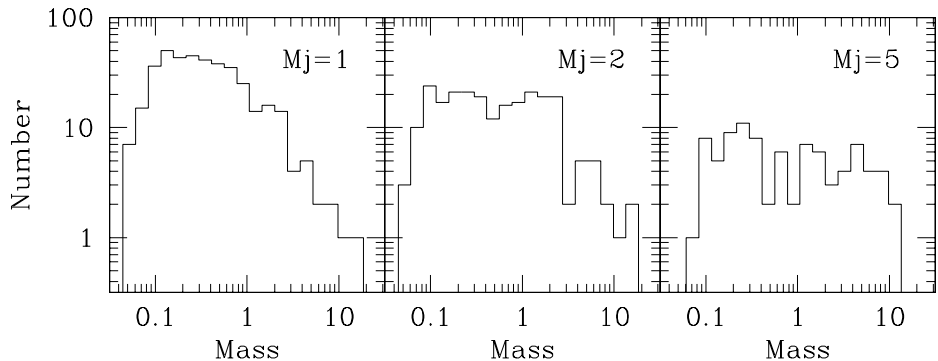,width=15.truecm,height=15.truecm,rwidth=15truecm,rheight=6.truecm}}
\caption{\label{IMFknee}  The mass function for the three isothermal simulations are shown at $t=1.5\tff$.
The knee of the IMF between the shallow slope at lower masses and the steeper slope at higher
masses scales with the Jeans mass of the simulation. 
The minimum mass resolved in these simulations is $m_{\rm lim} =  0.15 \solm$ such that any
downturn in the IMFs at low-masses is not resolved.}
\end{figure*}

\section{ Isothermal Cluster formation and stellar masses}
The early evolution of the  simulations is similar to that reported in Bonnell, Bate \& Vine (2003).
The initially uniform density is stirred up due to the chaotic motions present from the initial 'turbulent'
velocity field.  Shock compression leads to filamentary structures being formed very quickly, but these
do not proceed to form stars until larger, bound structures are formed through self-gravity (e.g. Clark \& Bonnell 2005).  Once gravitationally bound clumps form along the filaments, they collapse to form stars. These stars
fall towards their local potential minima and there form small clusters which grow by accreting stars and gas (e.g. Bonnell \etal~ 2003).  This process is illustrated in Figure~\ref{cluform} for simulation A. Note that
apart from a different random seed for the initial velocity field, the simulation is the same as reported in Bonnell \etal~ (2003).

The primary difference in the three isothermal simulations is the number of stars that form. Table 1 denotes the various parameters
of the three simulations, the number of stars formed by $t=1.5 \tff$,  and the associated total mass, as well as the average and maximum stellar mass. Simulation A, with a Jeans mass of $\mj =1\solm$, forms over 400 stars  by $t=1.5 \tff$ (over 500 by $t=1.72 \tff$) whereas
simulation B, with $\mj=2\solm$, forms just over 200 and simulation C, with $\mj=5\solm$, forms under
100 stars.  The lower density in the higher Jeans mass simulations means that the
clumps need to be more massive to overcome thermal (and kinetic) support to become gravitationally bound and collapse. Thus fewer of the clumps form stars and these stars are also more massive. 

In all three simulations, the dissipation of the supporting kinetic energies in shocks allows for
a larger scale gravitational collapse. This increases the gas density and thus lowers the Jeans mass
relative to its initial value. Thus stars that form later can have smaller initial masses. Gas accretion onto
all the stars increases their masses with those in the centre of clusters accreting more and therefore attaining the highest masses (Bonnell \etal~2001a; Bonnell \etal~2004).  This results in an increasing
dispersion in stellar masses with time and populates the resulting mass spectrum. Figure~\ref{massevol}
plots the evolution of the average and maximum stellar mass as a function of free-fall time for the
three simulations. The average stellar mass remains nearly constant for each simulation, and
scales (though not linearly) with the initial Jeans mass. In contrast, the maximum stellar mass
increases with time but does not depend on the initial Jeans mass. This is because the mass
of these stars is from gas accretion and not from the fragmentation process (Bonnell \etal~2001b; Bonnell \etal~2004). The total mass accreted by a star depends on the gas inflowing into the cluster's potential
well and not on the Jeans mass.

\subsection{Initial mass functions}

The resulting initial mass functions (IMFs) for the three isothermal simulations are shown in figure~\ref{IMFknee}.
The IMF for simulation A with $\mj=1\solm$ closely resembles that produced by simulations using similar innitial conditions
(Bonnell \etal~2003) as well as the observed IMF of young stellar clusters. The shallow slope
at low-masses extends to $0.5$ to 1 \solmas\  before developing into a steeper Salpeter-type
slope for higher masses. In contrast, the IMFs for simulations B and C show marked departures
in that the shallow part of the mass spectrum extends to significantly higher masses.
Simulation B, with  $\mj=2\solm$ has a shallow IMF extending to  $\approx 2$ or $3$ \solmas\ 
with a steeper Salpeter-like slope at higher masses. Simulation C,  with  $\mj=5\solm$ 
has the shallow part of the IMF extending to masses of $\approx 5$ to 10 \solmas, with only a slight indication of a steeper slope at higher masses.  It is clear from these IMFs that the knee or break-point between the shallow and steep sections of the IMF increases with the Jeans mass. To a fairly good approximation, the knee is located at the Jeans mass of the initial conditions.

The overall form of the IMF from simulation C resembles that found in the case of a low-mass
cloud with a Jeans mass of $\mj=1\solm$ and resolving down to the opacity limit (Bate \etal~2003). Both IMFs have a shallow
(flat in log mass) IMF which ends near the Jeans mass of the cloud. 
Similarly, when the Jeans mass in the lower-mass cloud is reduced, this break point
also moves to lower masses (Bate \& Bonnell 2005) as seen in Simulation B. This lends credence
to our conclusion that it is the Jeans mass which sets at which mass the IMF becomes steeper.

The physical interpretation behind this result is that clumps become bound near the initial
Jeans mass of the unperturbed cloud (Clark \& Bonnell 2005; Bonnell \etal~2004). These
clumps can subfragment as they are at least partially kinetically supported, such that
the thermal energy comprises only a fraction of the virial support in the clump. 
 Higher-mass stars require substantial accretion from beyond the
initial clump (Bonnell \etal~2004). The steep Salpeter-like slope appears once
competitive accretion in a stellar dominated regime occurs (Bonnell \etal~2001b),
and this can only happen once the stellar masses are above the masses of the
initial clumps, and hence the Jeans mass. 

\section{Non-isothermal cluster formation}

\begin{figure*}
\centerline{\psfig{figure=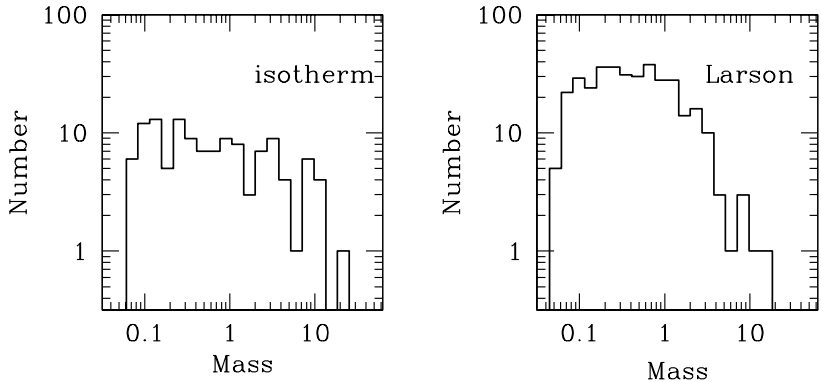,width=13.truecm,height=13.truecm,rwidth=13truecm,rheight=6.truecm}}
\caption{\label{IMFlariso}  The mass function for the two simulations starting with a  Jeans mass of $\mj = 5\solm$  are shown at $t=1.55\tff$. The isothermal simulation (simulation E, left panel) produces an unrealistically
flat IMF extending to high masses whereas the Larson (2005) type barotropic equation of state (simulation D, right panel)
produces a realistic IMF with a characteristic knee at $\approx 1 \solm$. This barotropic equation of state effectively reproduces the low Jeans mass results from more general initial
conditions.  
The minimum mass resolved in these simulations is $m_{\rm lim} =  0.15 \solm$ such that any
downturn in the IMFs at low-masses is not resolved.}
\end{figure*}

If the resulting IMFÊdepends on the initial Jeans mass, as these experiments suggest, then a universal IMF would appear to require fine tuning of the cloud properties such that the initial Jeans mass is always the same, with a value around a solar mass.
This is an uncomfortable conclusion, since, for a cloud of fixed
mass, the Jeans mass varies with both the temperature and cloud radius
to a high power ($1.5$). It would be far more reassuring if we could
instead invoke some physical process that yields a Jeans mass of
around a solar mass from a variety of initial cloud conditions.

One straightforward way
that this can occur is if the thermal physics results in an equation of state which 
has structure in it such that the temperature reaches a minimum at a certain density.
Based on the expected thermal physics of molecular clouds, Larson (2005) has recently suggested that the equation of state  of the form $P = k \rho^{\eta}$ ($T\propto \rho^{\eta -1}$) should include some deviation from isothermality
such that the gas cools ($\eta < 1$) with compression at low densities while at higher densities it heats ($\eta > 1$) slowly with increasing density.  While the gas is cooling, the  Jeans mass
actually decreases under compression at a greater relative rate  than does the local dynamical time 
whereas once the heating starts, the reverse occurs. The free-fall time denotes how quickly
an object can reconfigure itself due to self-gravity. For an isothermal gas, the free-fall time and Jeans
mass evolve in unison such that while the Jeans mass decreases, gravity induces a central
higher density region which contains exactly one Jeans mass. But, with cooling, a collapsing
object can develop multiple Jeans masses {\sl before} it can reconfigure
itself into a more stable, condensed configuration. 
Separate sub-entities can therefore develop, become gravitationally unstable and collapse
to form lower-mass objects. 
With heating under compression, the relative decrease in the Jeans mass is slowed such
that a  collapsing
clump can reconfigure itself to maintain one single, albeit more centrally condensed, object.
It then maintains the same object mass. 
The Jeans mass effectively gets hung up at the point where the equation of state
changes and this sets a mass scale
for star formation. 

In order to test this possibility, we performed one simulation with a barotropic equation of state
including both a cooling term at low densities and a heating term at higher densities. Starting
with an initial Jeans mass of $\mj=5$ \solmas\ , which in the isothermal runs gave a
flat IMF, we followed the collapse and fragmentation that forms a stellar cluster. This simulation
formed 334 stars within $t=1.5 \tff$, with a median and mean stellar mass of $0.47$ and $1.0$ \solmas,
respectively. The maximum stellar mass is $ 20$ \solmas. These values are very similar to
the isothermal $\mj=1$ \solmas\  simulation even though the initial Jeans mass was 5 times greater.

The corresponding IMF is  shown in Figure~\ref{IMFlariso} along with that from the corresponding isothermal simulation. The two simulations started
from the exact same initial conditions with a Jeans mass of $\mj = 5 \solm$ but produce
remarkably different IMFs. The isothermal simulation produces a flat IMF that extends from low masses 
up to $\approx 5 \solm$ as found above. In contrast, the barotropic equation of state produces
an IMF that has a broad peak that extends up to $\approx 1 \solm$ and a Salpeter-like slope
at higher masses. This observationally realistic IMF is similar to that produced when the initial Jeans mass is $\mj=1 \solm$. This is
due to the cooling which decreases the Jeans mass with gas compression. Thus, we can
confirm that the physically motivated barotropic equation of state suggested by Larson (2005) allows for more general initial conditions
to produce realistic IMFs.

A further point worth noting here is that this relaxation of the initial conditions also allows
for more flexibility in the resulting stellar densities in the newly formed clusters without changing the
location of the knee of the IMF.  The median stellar densities in simulation D reported here are a factor of ten smaller than
those found in Bonnell \etal (2003). This is most likely due to the larger initial Jeans radius
in this simulation as the gas density is lower  ($R_J \propto \rho^{-0.5}$). The Jeans
radius sets the scale for the fragmentation such that more widely spaced fragments
result in a less dense stellar cluster. 
\section{Conclusions}

Numerical simulations of the isothermal fragmentation of turbulent molecular clouds, and the subsequent competitive accretion 
in the forming stellar clusters, are able to reproduce the observed stellar IMF {\sl only}
if the initial Jeans mass is $\approx 1 \solm$. We find that the {\sl knee} of the IMF,
is approximately given by the Jeans mass. Thus, simulations with $\mj=5$ \solmas\ 
result in a shallow low-mass IMF that extends to $\simgreat 5 \solm$ in disagreement with
observations of the field and cluster star IMF. 

The main implication of this result is that in order to produce a universal IMF from a variety of cloud initial conditions, the
thermal physics in molecular clouds must result in a Jeans mass of order $1 \solm$ 
when star formation is initiated. A barotropic equation of state, involving a cooling term
at low densities followed by a gentle heating term once dust cooling dominates, is able
to produce a realistic IMF even when the initial Jeans mass is high ($\mj=5$ \solmas\ ).
We can conclude therefore that the small departures from isothermal  physics invoked by Larson (2005)
is sufficient to reproduce the characteristic stellar mass and thus realistic IMFs independent
of the exact initial conditions used (see also Jappsen \etal 2005). 
Variations in cooling physics,
at different epochs and in different metallicity regimes, would however
result in a shift of the location of the knee, implying a top heavy IMF
in regions of Population III star formation (Clarke \& Bromm 2003;  Bromm \& Larson 2004)

\section*{Acknowledgments}
We thanks the referee, Richard Larson, for useful comments which improved
the clarity of the text.
The computations reported here were performed using the U.K. Astrophysical
Fluids Facility (UKAFF). MRB is grateful for the support of a Philip Leverhulme Prize.

\end{document}